\documentclass[published]{JHEP3}
\JHEP{09(2003)015}

\usepackage{epsfig}

\skip\footins = 1\bigskipamount plus 2pt minus 4pt

\newcommand{\Gammait}{{\mit\Gamma}}
\newcommand{\tr}{\mathop{\rm tr}\nolimits}
\newcommand{\str}{\mathop{\rm str}\nolimits}
\newcommand{\Tr}{\mathop{\rm Tr}\nolimits}
\newcommand{\SU}{\mathop{\rm SU}\nolimits}
\newcommand{\U}{\mathop{\rm {}U}\nolimits}
\newcommand{\rmd}{{\rm d}}
\newcommand{\rmD}{{\rm D}}

\title{Wess-Zumino-Witten term on the lattice}

\author{Takanori Fujiwara and Hiroshi Suzuki\\
        Department of Mathematical Sciences, Ibaraki University\\ 
	Mito 310-8512, Japan\\
	E-mail: \email{fujiwara@mx.ibaraki.ac.jp},
                \email{hsuzuki@mx.ibaraki.ac.jp}}

\author{Kosuke Matsui and Masaru Yamamoto\\
        Graduate School of Science and Engineering, Ibaraki University\\ 
	Mito 310-8512, Japan\\
	E-mail: \email{matsui@serra.sci.ibaraki.ac.jp},
                \email{masaru@serra.sci.ibaraki.ac.jp}}

\abstract{We construct the Wess-Zumino-Witten (WZW) term in lattice
  gauge theory by using a Dirac operator which obeys the
  Ginsparg-Wilson relation.  Topological properties of the WZW term
  known in the continuum are reproduced on the lattice as a
  consequence of a non-trivial topological structure of the space of
  admissible lattice gauge fields. In the course of this analysis, we
  observe that the gauge anomaly generally implies that there is no
  basis of a Weyl fermion which leads to a single-valued expectation
  value in the fermion sector. The lattice Witten term, which carries
  information of a gauge path along which the gauge anomaly is
  integrated, is separated from the WZW term and the multivaluedness
  of the Witten term is shown to be related to the homotopy
  group~$\pi_{2n+1}(G)$. We also discuss the global ${\rm SU}(2)$ anomaly
  on the basis of the WZW term.  }

\received{August 1, 2003}
\accepted{September 4, 2003}

\begin{document} 

\section{Introduction}\label{section1}

In continuum gauge theories in $d=2n$ dimensions, topological
structures of the gauge theory and quantum anomalies are closely
related. The configuration space of gauge fields defined on~$S^{2n}$
is decomposed into an infinite number of connected components
classified by the homotopy group $\pi_{2n-1}(G)$, where $G$ is the
structure group of the gauge theory. By knowing the index of the Dirac
operator, one can see to which connected component the background
gauge field belongs. The quantum anomaly related to this topology of
the gauge field is the axial anomaly~\cite{Fujikawa:1979ay}. Although
each connected component~$\mathcal{A}$ of the configuration space,
being an affine space, is topologically trivial, the gauge orbit
space~$\mathcal{A}/\mathcal{G}$, where $\mathcal{G}$ is the group of
gauge transformations, generically has a non-trivial topology. In
particular, there may exist a non-contractible 2-sphere in the gauge
orbit space which gives rise to a
non-trivial~$\pi_2(\mathcal{A}/\mathcal{G})=\pi_1(\mathcal{G})$. The
latter implies the existence of a non-contractible loop in
$\mathcal{G}$ and, when~$G$ is simply-connected as we shall assume
below, it is identical to the homotopy
group~$\pi_{2n+1}(G)$. Physically, the existence of a non-contractible
2-sphere in the gauge orbit space gives rise to an obstruction in
defining a gauge invariant fermion determinant, i.e., the gauge
anomaly~\cite{Alvarez-Gaume:1983cs}.

In lattice gauge theory, any field configuration can smoothly be
deformed into one another and, hence, the space of lattice gauge
fields is intrinsically topologically trivial. However, it has been
shown~\cite{Luscher:1981zq} that by imposing the \emph{admissibility
  condition}
\begin{equation}
\|1-R[U(p)]\|<\epsilon,\qquad\hbox{for all plaquettes~$p$}\,,
\label{onexone}
\end{equation}
where $U(p)$ is the product of link variables around the plaquette~$p$
and $R$ stands for a unitary representation of the gauge group, it is
possible to endow the space of lattice gauge fields with a non-trivial
topological structure. The space of admissible lattice gauge fields is
decomposed into a (finite) number of connected components and one can
assign to each connected component~$\mathfrak{U}$ a topological
charge, which coincides with the Chern-Pontryagin index in the
classical continuum limit~\cite{Luscher:1981zq}.  For the overlap
Dirac operator~\cite{Neuberger:1998fp} which satisfies the
Ginsparg-Wilson (GW) relation~\cite{Ginsparg:1982bj}, the
admissibility condition is required to guarantee the regularity of the
Dirac operator~\cite{Hernandez:1999et}.\footnote{For the overlap Dirac
  operator, $R$ in eq.~(\ref{onexone}) is the gauge-group
  representation of the fermion and the Dirac operator is guaranteed
  to be regular if $\epsilon<(2-\sqrt{2})/[2n(2n-1)]$.} Then the index
of the Dirac operator, whose density is the axial anomaly on the
lattice, is constant over~$\mathfrak{U}$ due to the lattice index
theorem~\cite{Hasenfratz:1998ri,Neuberger:1998fp}. It is expected that
the topological charge of ref.~\cite{Luscher:1981zq} and the index are
identical when $\epsilon$ is sufficiently
small~\cite{Narayanan:1997sa}.

A removal of non-admissible configurations from the space of gauge
fields may leave ``holes'' in a connected component~$\mathfrak{U}$. If
this occurs, one may have non-trivial $\pi_2(\mathfrak{U})$ and this
could cause, as we will see below, an obstruction in defining a smooth
integration measure for Weyl fermions.

The basic picture behind a classification of topological sectors in
lattice gauge theory however appears quite different from that in the
continuum gauge theory. In the latter, the gauge potential approaches
to a pure gauge configuration on the $2n-1$ dimensional sphere at
infinity and this defines a mapping from the $2n-1$ sphere to~$G$. The
space of gauge potentials is thus classified by~$\pi_{2n-1}(G)$. In
lattice gauge theory, on the other hand, the
admissibility~(\ref{onexone}) is a gauge invariant condition and does
not place any restriction on gauge degrees of freedom. The space of
lattice gauge transformations~$\mathfrak{G}$ is topologically trivial
even after imposing eq.~(\ref{onexone}). Hence the idea of classifying
topological sectors of the gauge field by $\pi_{2n-1}(G)$ cannot be
applied to lattice gauge theory as it stands.  Similarly, on the
lattice $\pi_1(\mathfrak{G})=0$ and the connection between a
nontrivial $\pi_1(\mathcal{G})=\pi_{2n+1}(G)$ and the gauge anomaly
seems to be lost in lattice gauge theory. This makes the construction
of the Wess-Zumino-Witten (WZW) term~\cite{Wess:yu,Witten:tw} in
lattice gauge theory intricate because the multivalued nature of the
WZW term in the continuum theory relies on the non-triviality
of~$\pi_{2n+1}(G)$.

In this paper, on the basis of a Dirac operator which obeys the GW
relation, we formulate the WZW term on the lattice. Our WZW term is
manifestly local on the lattice and has a correct classical continuum
limit. Our construction can be regarded as an ``anomalous version'' of
the formulation of refs.~\cite{Luscher:1999du,Luscher:1999un} in which
anomaly-free chiral gauge theories are studied. Our task is relatively
simple compared to that for anomaly-free
cases~\cite{Luscher:1999du,Luscher:1999un} because we treat anomalous
cases and it is not necessary to worry about the gauge anomaly
cancellation on the lattice, which is a highly non-trivial issue.
Unfortunately, however, our construction in its explicit form is
applied only to a restricted class of gauge fields and gauge
transformations and it may not be suitable for numerical
simulations. Nevertheless, by using our construction, we can
understand how various properties of the WZW term expected in the
continuum on topological grounds are realized in the lattice
framework, despite the above mentioned differences between the
topology of gauge fields in continuum and lattice frameworks.

\looseness=1 Our motivation is not new, and in fact, many related works can be
found in the literature. Ref.~\cite{Aoki:1985hb} studied a
construction of the WZW term in the context of Wilson fermion. The
overlap formulation for lattice chiral gauge
theories~\cite{Narayanan:1993wx,Narayanan:1995gw} is closely related
to the formulation of refs.~\cite{Luscher:1999du,Luscher:1999un}. An
implementation of the WZW term by the overlap has been proposed in
ref.~\cite{Narayanan:1995gw}.\footnote{However the locality is not
  obvious in this implementation.}  With the overlap formulation, the
WZW term in 2-dimensions has been studied analytically and
numerically~\cite{Kikukawa:1996ur}. To our knowledge, however,
topological aspects of the WZW term on the lattice have not fully been
investigated so far.  In a later section, we will see that our WZW
term on the lattice reproduces the winding number of the phase of
chiral determinant along a gauge
loop~\cite{Alvarez-Gaume:1983cs}. This is basically the result
obtained in ref.~\cite{Adams:2000yi} (in ref.~\cite{Adams:1999um}, it
was shown that the winding number can be expressed by the index of a
Dirac operator in $2n+2$ dimensions). We will observe that a
non-trivial winding number of chiral determinant, i.e, the gauge
anomaly, is a consequence of non-contractible 2-spheres
in~$\mathfrak{U}$, which is classified by~$\pi_{2n+1}(G)$. An
existence of such non-contractible surfaces in~$\mathfrak{U}$ has also
been pointed out~\cite{Adams:2002ms}. In the present paper, we show
that a non-trivial winding number directly implies an obstruction in
defining a smooth fermion measure, not only an obstruction for the
gauge invariance; anomalous chiral gauge theories in general cannot
consistently be formulated in the framework of
refs.~\cite{Luscher:1999du,Luscher:1999un}. Our way of
argument\footnote{In ref.~\cite{Kikukawa:2002ms}, by using a similar
  argument, a non-contractible 2-torus in~$\mathfrak{U}$ of abelian
  theories was identified as an obstruction for a smooth fermion
  measure. According to ref.~\cite{Adams:2002tb}, the argument of
  ref.~\cite{Adams:2002ms} shows that an obstruction in the orbit
  space~$\mathfrak{U}/\mathfrak{G}$ of abelian theories observed in
  ref.~\cite{Neuberger:1998xn} implies a non-contractible 2-torus
  in~$\mathfrak{U}$.} to show this fact is simple and also provides a
coherent picture to understand topological aspects of the WZW term.

This paper is organized as follows. Section~\ref{section2} is devoted
to a construction of the WZW term in lattice gauge theory, based on
the Ginsparg-Wilson relation. Then, in section~\ref{section3}, we show
that the WZW term exhibits an ambiguity of $2\pi$ multiple of an
integer, depending on a choice of the gauge path along which the
lattice gauge anomaly is integrated. We also show that this ambiguity
can be separated as a term which does not depends on the gauge field
(a lattice analogue of the Witten term~\cite{Witten:tw}). In
section~\ref{section4}, we study the winding number of the chiral
determinant~\cite{Alvarez-Gaume:1983cs} and show that it is directly
related to an obstruction to a smooth fermion measure. This shows that
anomalous chiral gauge theories are intrinsically different from
anomaly-free cases. In section~\ref{section5}, we show that the
multivaluedness of the Witten term~\cite{Witten:tw} is precisely
realized on the lattice and its value is in fact classified
by~$\pi_{2n+1}(G)$. In section~\ref{section6}, on the basis of the
result of~ref.~\cite{Bar:1999ka}, we illustrate how the WZW term
reproduces the global $\SU(2)$
anomaly~\cite{Witten:fp}. Section~\ref{section7} is devoted to
conclusion. In appendix~\ref{section8}, we collect relevant materials
for the classical continuum limit of the WZW term.

We consider $d=2n$ dimensional periodic lattice with a finite
size. The lattice spacing will be denoted by~$a$.

\section{Construction of the WZW term}\label{section2}

\subsection{Basic idea}\label{section2.1}

We first sketch our basic idea to define the WZW term on the
lattice. We start with the chiral determinant of a (left-handed) Weyl
fermion on the lattice
\begin{equation}
\det M[U]=\int\rmD[\psi]\rmD[\overline\psi]\,e^{-S_{\rm F}}\,,\qquad
S_{\rm F}=a^{2n}\sum_x\overline\psi(x)D\psi(x)\,,
\label{twoxone}
\end{equation}
where $U$ denotes the lattice gauge field (link variables). In most
part of this paper, we assume that the gauge field~$U$ belongs to the
vacuum sector. Namely, we assume that $U$ can smoothly be deformed
into the trivial configuration~$U=1$ within the space of admissible
gauge fields specified by eq.~(\ref{onexone}); in eq.~(\ref{onexone})
$R$ stands for the gauge-group representation of the fermion.

In eq.~(\ref{twoxone}), the Dirac operator~$D$ is assumed to satisfy
the GW relation
\begin{equation}
   \gamma_5D+D\gamma_5=aD\gamma_5D\,.
\label{twoxtwo}
\end{equation}
Although we will not need the explicit form of~$D$ in this paper, we
assume following properties of~$D$: The $\gamma_5$-hermiticity
$D^\dagger=\gamma_5D\gamma_5$, the gauge covariance, the
locality\footnote{For the precise definition of the locality assumed
  here, see ref.~\cite{Luscher:1999du}.} and absence of species
doublers. These properties would imply the Dirac operator reproduces
the correct axial anomaly in the classical continuum
limit~\cite{Kikukawa:1998pd,Fujiwara:2002xh}; we also assume
this. Neuberger's overlap Dirac operator~\cite{Neuberger:1998fp}
possesses all these properties if configurations of gauge field are
restricted by the
condition~(\ref{onexone})~\cite{Hernandez:1999et}. Following
refs.~\cite{Narayanan:1998uu}, we then introduce the modified chiral
matrix~$\hat\gamma_5=\gamma_5(1-aD)$, which fulfills
\begin{equation}
   (\hat\gamma_5)^\dagger=\hat\gamma_5,\qquad(\hat\gamma_5)^2=1\,,\qquad
D\hat\gamma_5=-\gamma_5D
\label{twoxthree}
\end{equation}
and define chiral projectors by
\begin{equation}
   \hat P_\pm={1\over2}(1\pm\hat\gamma_5)\,,\qquad
   P_\pm={1\over2}(1\pm\gamma_5)\,,\qquad D\hat P_\pm=P_\mp D.
\label{twoxfour}
\end{equation}
Then in eq.~(\ref{twoxone}), the left-handed Weyl fermion is defined
by imposing the constraints
\begin{equation}
   \hat P_-\psi=\psi,\qquad\overline\psi=\overline\psi P_+\,.
\label{twoxfive}
\end{equation}
In this construction, the chirality is consistently imposed due to the
last relation of~eq.~(\ref{twoxfour}). The action in
eq.~(\ref{twoxone}) is manifestly gauge invariant and consistent with
the locality. We postpone the detailed account on the fermion measure
$\rmD[\psi]\rmD[\overline\psi]$ of eq.~(\ref{twoxone}) to the next
subsection.

\pagebreak[3]

Assuming that the chiral determinant has been defined, we define the
WZW term on the lattice by
\begin{equation}
   i\Gammait_{\rm WZW}[g,U]=\ln\det M[U]-\ln\det M[U^{g^{-1}}]\,,
\label{twoxsix}
\end{equation}
where $U^g$ denotes the gauge transformation of the gauge field~$U$
parametrized by~$g$,
\begin{equation}
   U^g(x,\mu)=g(x)U(x,\mu)g(x+a\hat\mu)^{-1}\,.
\label{twoxseven}
\end{equation}
{}From its definition, it is obvious that the WZW term vanishes if the
chiral determinant is gauge invariant; it thus picks up only the
effect of gauge anomaly. Under the infinitesimal gauge
transformation,\footnote{$\nabla_\mu$ is the gauge covariant forward
  difference operator, $\nabla_\mu\omega(x)=
  [U(x,\mu)\omega(x+a\hat\mu)U(x,\mu)^{-1}-\omega(x)]/a$.}
\begin{eqnarray}
   \delta U(x,\mu)&=&\omega(x)U(x,\mu)-U(x,\mu)\omega(x+a\hat\mu)
   =-a\nabla_\mu\omega(x)U(x,\mu)\,,
\nonumber\\
   \delta g(x)&=&\omega(x)g(x)\,,
\label{twoxeight}
\end{eqnarray}
the combination~$U^{g^{-1}}$ in the second term of eq.~(\ref{twoxsix})
does not change and, on the other hand, the first term produces the
gauge anomaly. By this way, the WZW term, which will turn to be a
local functional on the lattice, reproduces the gauge anomaly as
$i\delta\Gammait_{\rm WZW}[g,U]$ $=\delta\ln\det M[U]$. In the continuum
theory, this is the defining relation of the WZW term.

\subsection{Fermion integration measure and the associated $\U(1)$ bundle}\label{section2.2}

To define the fermion integration measure in eq.~(\ref{twoxone}), we
introduce an orthonormal complete set of vectors in the constrained
space~(\ref{twoxfive}),
\begin{equation}
   \hat P_-v_j=v_j\,,\qquad (v_k,v_j)=\delta_{kj}\,.
\label{twoxnine}
\end{equation}
Note that $\sum_jv_j(x)v_j(y)^\dagger=\hat P_-(x,y)$. By expanding the
fermion field in terms of this basis
\begin{equation}
   \psi(x)=\sum_jv_j(x)c_j\,,
\label{twoxten}
\end{equation}
the integration measure is defined by
\begin{equation}
   \rmD[\psi]=\prod_j\rmd c_j\,.
\label{twoxeleven}
\end{equation}
Similarly, for the anti-fermion, by using a basis such that $\overline
v_kP_+=\overline v_k$, we set
\begin{equation}
   \rmD[\overline\psi]=\prod_k\rmd\overline c_k\,,\qquad
   \overline\psi(x)=\sum_k\overline c_k\overline v_k(x)\,.
\label{twoxtwelve}
\end{equation}
In terms of these bases, the chiral determinant~(\ref{twoxone}) is
given by the determinant of the matrix
\begin{equation}
   M_{kj}=a^{2n}\sum_x\overline v_k(x)Dv_j(x)\,.
\label{twoxthirteen}
\end{equation}

The above (seemingly simple) construction, however, poses complicated
problems.  First, the constraint~(\ref{twoxnine}) does not specify the
basis vectors uniquely. A different choice of basis, as we will see
shortly, leads to a difference in the phase of the fermion integration
measure. This phase moreover may depend on the gauge field and hence
may influence on the physical contents of the system. Secondly, a
connected component of the space of admissible gauge
fields~$\mathfrak{U}$ can be topologically non-trivial as we already
noted.  Thus it is not obvious whether there exist bases
over~$\mathfrak{U}$, with which the chiral determinant (or more
generally an expectation value of operators in the fermion sector) is
a single-valued function over~$\mathfrak{U}$. These closely related
problems can be formulated as
follows~\cite{Luscher:1999du,Luscher:1999un}.

We cover~$\mathfrak{U}$ by local coordinate patches~$X_A$, labeled
by~$A$.  Suppose that some basis~$v_j^A$ has been chosen within each
patch, as is always possible for contractible local
patches. Generally, however, on the intersection $X_A\cap X_B$ of two
patches, basis vectors $v_j^A$ and~$v_j^B$ are different and are
related by a unitary transformation
\begin{equation}
   v_j^B(x)=\sum_k v_k^A(x)\tau(A\to B)_{kj}\,.
\label{twoxfourteen}
\end{equation}
By definition, the transition function~$\tau(A\to B)$ satisfies the
cocycle condition
\begin{equation}
   \tau(A\to C)=\tau(A\to B)\tau(B\to C)\qquad\hbox{on $X_A\cap
     X_B\cap X_C$}
\label{twoxfifteen}
\end{equation}
and thus defines a unitary principal bundle over~$\mathfrak{U}$.
Corresponding to eq.~(\ref{twoxfourteen}), the fermion integration
measures defined in~$X_A$ and in~$X_B$ are related as
\begin{equation}
   \rmD[\psi]_B=\rmD[\psi]_Ag_{AB},\qquad g_{AB}=\det\tau(A\to
   B)\in\U(1)\,.
\label{twoxsixteen}
\end{equation}
This phase factor~$g_{AB}$ thus defines a $\U(1)$ bundle associated to
the fermion integration measure. For a sensible formulation of Weyl
fermions, various expectation values in the fermion sector must be a
single-valued function over~$\mathfrak{U}$. To realize this, one
re-defines basis vectors in each patch\footnote{Under a change of
  basis, the phase factor transforms as $g_{AB}\to h_Ag_{AB}h_B^{-1}$,
  where $h_A$ is the determinant of the transformation matrix between
  two bases in~$X_A$.} such that $g_{AB}=1$ on all
intersections. However, this is possible if and only if the $\U(1)$
bundle is trivial.

To find a characterization of this $\U(1)$ bundle, we consider a
variation of gauge field
\begin{equation}
   \delta_\eta U(x,\mu)=a\eta_\mu(x)U(x,\mu)\,,\qquad
   \eta_\mu(x)=\eta_\mu^a(x)T^a
\label{twoxseventeen}
\end{equation}
and define the \emph{measure term\/} within a patch, say $X_A$, by
\begin{equation}
   \mathfrak{L}_\eta^A=i\sum_j(v_j^A,\delta_\eta v_j^A)\,.
\label{twoxeighteen}
\end{equation}
This characterizes how the basis vectors change under the
variation. On the intersection, $X_A\cap X_B$, we see from
eq.~(\ref{twoxfourteen}) that the measure terms in $X_A$ and in~$X_B$
are related as
\begin{equation}
   \mathfrak{L}_\eta^B=\mathfrak{L}_\eta^A+ig_{AB}^{-1}\delta_\eta
   g_{AB}\,.
\label{twoxnineteen}
\end{equation}
This shows that the measure term is the $\U(1)$ connection associated
to the $\U(1)$ bundle. By applying the identity\footnote{Here we have
  assumed that the variations $\eta$ and $\zeta$ are independent of
  the gauge field.}
\begin{equation}
   \delta_\eta\delta_\zeta-\delta_\zeta\delta_\eta
   +a\delta_{[\eta,\zeta]}=0
\label{twoxtwenty}
\end{equation}
to the definition of the measure term~(\ref{twoxeighteen}), we find
\begin{equation}
   \delta_\eta\mathfrak{L}_\zeta^A-\delta_\zeta\mathfrak{L}_\eta^A
   +a\mathfrak{L}_{[\eta,\zeta]}^A =i\Tr(\hat P_-[\delta_\eta\hat
     P_-,\delta_\zeta\hat P_-])\,.
\label{twoxtwentyone}
\end{equation}
The right hand side, which depends neither on the patch label nor on
basis vectors, is nothing but the curvature of the $\U(1)$
bundle. Then as a measure of non-triviality of the $\U(1)$ bundle, we
may consider a 2-dimensional integral of the curvature, the first
Chern number
\begin{equation}
   {1\over2\pi}\int_{\mathcal{M}}\rmd t\,\rmd s\,
   i\Tr(P_{t,s}[\partial_tP_{t,s},\partial_sP_{t,s}])\,,
\label{twoxtwentytwo}
\end{equation}
which is an \emph{integer}. $\mathcal{M}$ in this expression is a
\emph{closed\/} 2-surface in~$\mathfrak{U}$ and $t$ and~$s$ are local
coordinates of~$\mathcal{M}$. Namely, we defined two-parameter family
of gauge fields $U_{t,s}$ in~$\mathfrak{U}$, and accordingly defined
the projection operators by~$P_{t,s}=\hat P_-|_{U=U_{t,s}}$. If the
above integer does not vanish, the $\U(1)$ bundle is non-trivial and
there exists no choice of basis vectors such that expectation values
in the fermion sector are single-valued over~$\mathcal{M}$. Namely,
the Weyl fermion cannot consistently be formulated.  In other words,
$\mathcal{M}$ is a non-contractible 2-surface in~$\mathfrak{U}$ which
gives rise to a topological obstruction in defining a smooth fermion
measure.

In the above argument, we started with basis vectors defined patch by
patch.  This way of argument, however, is not convenient in
constructing the fermion measure which ensures the locality. The
measure term~$\mathcal{L}_\eta$, being linear in the variation~$\eta$,
can be expressed as
\begin{equation}
   \mathcal{L}_\eta=a^{2n}\sum_x\eta_\mu^a(x)j_\mu^a(x)\,,
\label{twoxtwentythree}
\end{equation}
where $j_\mu^a(x)$ is termed the \emph{measure current}. It can be
shown~\cite{Luscher:1999du} that the expected locality of the system
is guaranteed if the measure current is a local function of link
variables. To ensure the correct physical contents of the formulation,
therefore, it is convenient to start with a certain local measure
current. Then according to the reconstruction
theorem~\cite{Luscher:1999du,Luscher:1999un}, if the local current
satisfies the following conditions, one can re-construct basis vectors
which lead to a smooth fermion measure that is consistent with the
locality.  The measure term~(\ref{twoxeighteen}) of the so constructed
basis coincides with the measure term defined from the current by
eq.~(\ref{twoxtwentythree}).  (1)~The current is a smooth (i.e.,
single-valued) function of the gauge fields contained
in~$\mathfrak{U}$. (2)~The measure term~(\ref{twoxtwentythree})
defined from the current satisfies the \emph{global integrability}
\begin{equation}
   \exp\left(i\int_0^1\rmd t\,\mathcal{L}_\eta\right)
   =\det(1-P_0+P_0Q_1)
\label{twoxtwentyfour}
\end{equation}
along any \emph{closed loop\/} in~$\mathfrak{U}$. In this expression,
the loop~$U_t$ ($U_1=U_0$) is parametrized by~$t\in[0,1]$ and the
variation is given by
$a\eta_\mu(x)=\partial_tU_t(x,\mu)U_t(x,\mu)^{-1}$. Projection
operators along the loop are defined by~$P_t=\hat P_-|_{U=U_t}$. The
operator~$Q_t$ is defined by the differential equation~$\partial_t
Q_t=[\partial_tP_t,P_t]Q_t$ and~$Q_0=1$. Thus the task to construct
the fermion measure is reduced to find an appropriate local
current~$j_\mu^a(x)$ which satisfies the above two
conditions.\footnote{For a gauge invariant formulation of anomaly-free
  cases, an additional condition, the anomaly cancellation on the
  lattice~\cite{Luscher:1999du,Luscher:1999un}, has to be fulfilled.}
For an infinitesimally small (contractible) loop in $\mathfrak{U}$, it
can be shown that the global integrability becomes the \emph{local
  integrability\/}
\begin{equation}
   \delta_\eta\mathfrak{L}_\zeta-\delta_\zeta\mathfrak{L}_\eta
   +a\mathfrak{L}_{[\eta,\zeta]} =i\Tr(\hat P_-[\delta_\eta\hat
     P_-,\delta_\zeta\hat P_-])
\label{twoxtwentyfive}
\end{equation}
that is nothing but eq.~(\ref{twoxtwentyone}) without patch labels
(because here we started with a measure current defined globally
over~$\mathfrak{U}$).  In this paper, we adopt this latter viewpoint
which starts with the measure current and seek for an appropriate
local measure current, or equivalently the measure term.

\subsection{WZW term with the GW fermion}\label{section2.3}

{}From the expression~(\ref{twoxthirteen}), we see that the
infinitesimal variation of the chiral determinant is given by
\begin{equation}
   \delta_\eta\ln\det M[U]=\Tr(\delta_\eta D\hat P_-D^{-1}P_+)
   -i\mathfrak{L}_\eta\,.
\label{twoxtwentysix}
\end{equation}
Therefore the WZW term~(\ref{twoxsix}) can be obtained by integrating
this variation along a sequence of gauge transformations (a gauge
path). We introduce a one-parameter family of gauge transformations
\begin{equation}
   g_t(x),\qquad 0\leq t\leq 1
\label{twoxtwentyseven}
\end{equation}
such that
\begin{equation}
   g_0(x)=1,\qquad g_1(x)=g(x)\,.
\label{twoxtwentyeight}
\end{equation}
\EPSFIGURE[t]{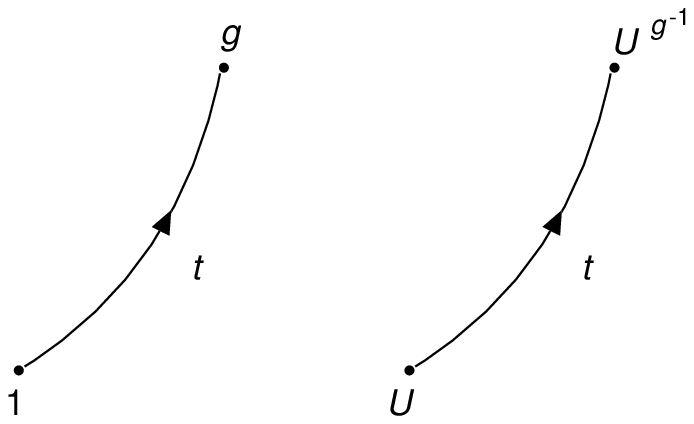,width=.6\textwidth}{The paths in the space of gauge transformations
(left) and in the space of gauge field configurations (right) along which we
integrate a gauge variation of the chiral determinant to define the WZW term.\label{fig1}}
See figure~\ref{fig1}. An example of such a path is given by
\begin{equation}
   g_t(x)=[g(x)]^t.
\label{twoxtwentynine}
\end{equation}
In lattice gauge theory, such a smooth path~$g_t$ which connects $1$
and~$g$ \emph{always\/} exists because $\pi_0(\mathfrak{G})=0$; the
situation is quite different from the continuum. Corresponding to this
path, we define a one-parameter family of gauge fields by
\begin{equation}
   U_t(x,\mu)=U^{g_t^{-1}}(x,\mu),\qquad 0\leq t\leq 1\,,
\label{twoxthirty}
\end{equation}
which connects $U_0=U$ and~$U_1=U^{g^{-1}}$ (figure~\ref{fig1}). Note
that, since the condition~(\ref{onexone}) is gauge invariant, if the
starting configuration~$U_0=U$ belongs to the space of admissible
gauge fields~$\mathfrak{U}$, then the path is also
within~$\mathfrak{U}$. By this way, we have
\begin{eqnarray}
   \Gammait_{\rm WZW}[g,U]
   &=&i\int_0^1\rmd t\,\partial_t\ln\det M[U^{g_t^{-1}}]
\nonumber\\
   &=&i\int_0^1\rmd t\Tr(\partial_t D\hat P_-D^{-1}P_+)
   +\int_0^1\rmd t\,\mathfrak{L}_\eta\,,
\label{twoxthirtyone}
\end{eqnarray}
where the variation in the second term is given by the gauge variation
along the gauge path
\begin{equation}
   \eta_\mu(x)=\nabla_\mu(g_t^{-1}\partial_t g_t)(x)\,.
\label{twoxthirtytwo}
\end{equation}

The first term of eq.~(\ref{twoxthirtyone}) can be rendered manifestly
local.  Noting the gauge covariance of the Dirac operator
\begin{equation}
   D|_{U=U_t}=R(g_t^{-1})D|_{U=U_0}R(g_t)
\label{twoxthirtythree}
\end{equation}
and the last property of eq.~(\ref{twoxfour}), we have
\begin{equation}
   \Gammait_{\rm WZW}[g,U] =a^{2n}\sum_x\mathcal{A}^a(x)|_U
   \int_0^1\rmd t\,(\partial_t g_tg_t^{-1})^a(x) +\int_0^1\rmd
   t\,\mathfrak{L}_\eta\,,
\label{twoxthirtyfour}
\end{equation}
where the \emph{covariant gauge anomaly\/} $\mathcal{A}^a(x)$ is
defined by
\begin{equation}
   \mathcal{A}^a(x)={ia\over2}\tr[\gamma_5R(T^a)D(x,x)]\,.
\label{twoxthirtyfive}
\end{equation}
This is a local function of gauge fields, because it does not contain
the inverse of the Dirac operator which is generically a non-local
combination of link variables.

\EPSFIGURE[t]{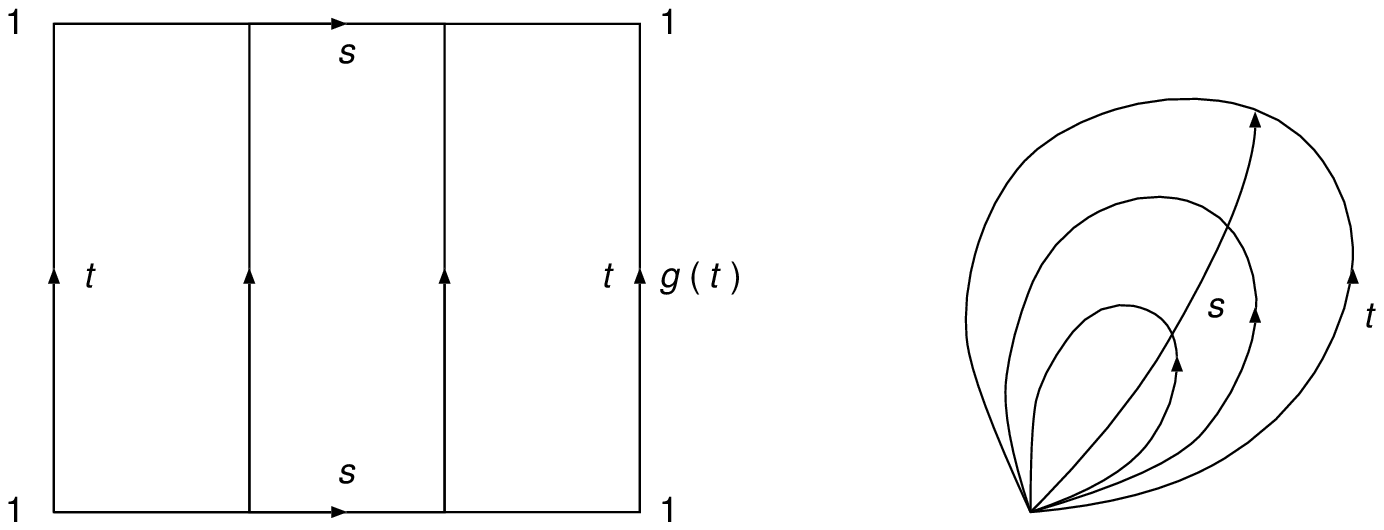}{The two-parameter family of gauge transformations (left)
and the two-parameter family gauge field configurations (right) utilized in
eq.~(\ref{twoxfortytwo}).\label{fig2}}

Now, the covariant gauge anomaly and the curvature in
eq.~(\ref{twoxtwentyfive}) are not independent. To see this, take the
gauge variation~$\eta_\mu(x)=-\nabla_\mu\omega(x)$ as one variation in
the curvature. Noting the gauge covariance $\delta_\eta\hat
P_-=[R(\omega),\hat P_-]$ and the identity~$\hat P_-\delta_\zeta \hat
P_-\hat P_-=0$, we find
\begin{equation}
   i\Tr(\hat P_-[\delta_\eta\hat P_-,\delta_\zeta\hat P_-])
   =-a^{2n}\sum_x\omega^a(x)\delta_\zeta\mathcal{A}^a(x)\,.
\label{twoxthirtysix}
\end{equation}
Namely, the curvature component in the direction of the gauge
variation is given by a variation of the covariant anomaly. This
relation is especially useful to find the classical continuum limit of
the curvature, as we illustrate in appendix~\ref{section8}. This
relation also has an important consequence in our discussions
below. Let us consider the case that the path~(\ref{twoxtwentyseven})
forms a \emph{closed loop}, namely $g_1=g_0=1$. For this case, it is
always possible to find a two-parameter family~$g_{t,s}$ of gauge
transformations such that
\begin{equation}
   g_{0,s}(x)=g_{t,0}(x)=1\,,\qquad 0\leq t\leq1,\quad0\leq s\leq1\,,
\label{twoxthirtyseven}
\end{equation}
and
\begin{equation}
   g_{t,1}(x)=g_t(x),\qquad0\leq t\leq1\,.
\label{twoxthirtyeight}
\end{equation}
See figure~\ref{fig2}. The example is
\begin{equation}
   g_{t,s}(x)=[g_t(x)]^s\,.
\label{twoxthirtynine}
\end{equation}
Since these generate gauge transformations, the corresponding
two-parameter family of gauge fields
\begin{equation}
   U_{t,s}(x,\mu)=U^{g_{t,s}^{-1}}(x,\mu)\,, \qquad 0\leq t\leq
   1,\quad0\leq s\leq1\,,
\label{twoxforty}
\end{equation}
is within the space of admissible gauge fields~$\mathfrak{U}$. Now we
integrate the relation~(\ref{twoxthirtysix}) over the disk represented
in figure~\ref{fig2}, by identifying $\eta$ as the gauge variation
along the $t$-direction,
\begin{equation}
   \eta_\mu(x)=\nabla_\mu(g_{t,s}^{-1}\partial_tg_{t,s})(x)
\label{twoxfortyone}
\end{equation}
and~$\zeta$ as the direction parametrized by~$s$. Noting the gauge
covariance of the covariant anomaly, we have
\begin{eqnarray}
   \int_0^1\rmd t\,a^{2n}\sum_x(g_t^{-1}\partial_t g_t)^a(x)
   \mathcal{A}^a(x)|_{U=U_t}
   &=&a^{2n}\sum_x\mathcal{A}^a(x)|_{U=U_0}
   \int_0^1\rmd t\,(\partial_t g_tg_t^{-1})^a(x)
\nonumber\\
   &=&i\int_0^1\rmd t\int_0^1\rmd s\,
   \Tr(P_{t,s}[\partial_tP_{t,s},\partial_sP_{t,s}])\,,
\label{twoxfortytwo}
\end{eqnarray}
where we have used the boundary conditions,
eqs.~(\ref{twoxthirtyseven}) and~(\ref{twoxthirtyeight}). We have
observed that, when the gauge path~(\ref{twoxthirty}) is a closed loop
in~$\mathfrak{U}$, the first term of the WZW
term~(\ref{twoxthirtyfour}) (a contribution of the covariant anomaly),
can be expressed as the surface integral of the curvature over the
disk~(\ref{twoxforty}) (this always exists within~$\mathfrak{U}$)
which is spanned by the loop. This property will be repeatedly
utilized in the following discussions.

\FIGURE[t]{\centerline{\epsfig{file=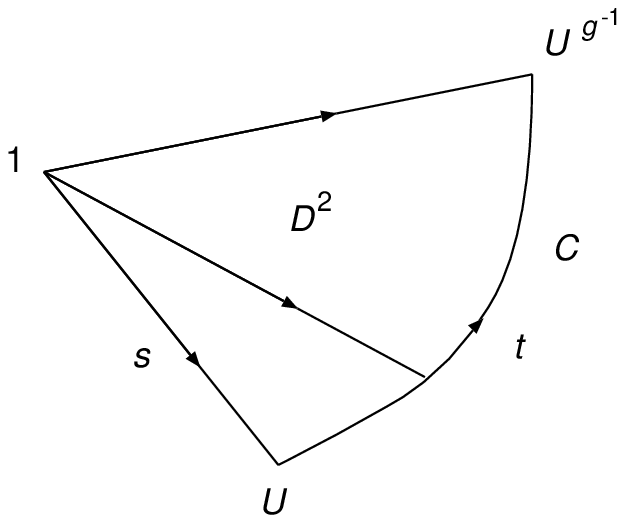,width=7.5cm}}%
\caption{The two-parameter family of gauge field
configurations to define the WZW term. The direction of~$s$ is to construct the
measure term~(\ref{twoxfortyeight}) and the gauge anomaly is integrated along
the gauge path~$C$, the direction of~$t$.\label{fig3}}}

\subsection{An explicit construction of the measure term}\label{section2.4}

It is obvious that the above definition of the WZW
term~(\ref{twoxthirtyfour}) is incomplete unless we provide an
explicit form of the measure term~$\mathfrak{L}_\eta$. To construct
the measure term, we again consider a two-parameter family of gauge
fields~$U_{t,s}$, such that
\begin{equation}
   U_{t,0}=1,\qquad 0\leq t\leq1\,,
\label{twoxfortythree}
\end{equation}
and
\begin{equation}
   U_{t,1}=U^{g_t^{-1}}\,,\qquad 0\leq t\leq1\,.
\label{twoxfortyfour}
\end{equation}
See figure~\ref{fig3}. Here we encounter a trouble. There is no
general guarantee that this family of gauge fields is contained in the
space of admissible gauge fields, $\mathfrak{U}$. Recall that we have
assumed that $U=U_{0,1}$ belongs to the vacuum sector. So the
existence of a smooth path $U_{0,s}$ which connects $1$ and~$U$ is
guaranteed by assumption. However, the admissibility of
general~$U_{t,s}$ is not. In fact, in a later section, we will
encounter an example in which such a smooth family does not
exist. Thus to ensure the admissibility, we place following
restrictions on $U$ and~$g$; these limit an applicability of our
construction. We assume that the gauge field~$U$ is given by a lattice
transcription of a certain gauge potential defined on a continuum
torus~$T^{2n}$.\footnote{The continuum limit of a periodic lattice
  is~$T^{2n}$. However, in connection with the continuum theory, we
  are interested in gauge fields defined on~$S^{2n}$. Hence, in what
  follows, we consider gauge fields on~$T^{2n}$ that can be regarded
  as those on~$S^{2n}$.  These can be obtained by using a surjection
  $T^{2n}\to S^{2n}$ whose degree of mapping is unity. Gauge fields
  and gauge transformations on~$T^{2n}$ are given by the pullback of
  those on~$S^{2n}$.} To be precise, link variables are given by the
path-ordered exponential of a smooth gauge potential
\begin{equation}
   U(x,\mu)=\mathcal{P} \exp\left[a\int_0^1\rmd
     u\,A_\mu(x+(1-u)a\hat\mu)\right].
\label{twoxfortyfive}
\end{equation}
Also we assume that the gauge transformation parameter~$g(x)$ is given
by a lattice restriction of a certain smooth $G$-valued function
defined on the torus. Under these restrictions, when the lattice
spacing is sufficiently small, there always exists a two-parameter
family of lattice gauge fields in~$\mathcal{M}$ which possesses the
above boundary values. For example, a choice
\begin{equation}
   U_{t,s}(x,\mu)=[g_t^{-1}(x)U(x,\mu)g_t(x+a\hat\mu)]^s
\label{twoxfortysix}
\end{equation}
does the job.

\pagebreak[3]

With the above restrictions, it is easy to find an appropriate measure
term.  One way is to introduce an additional parameter~$r$ and set
$a\eta_\mu(x)=\partial_t U_{t,s,r}(x,\mu)U(x,\mu)_{t,s,r}^{-1}$ and
$a\zeta_\mu(x)=\partial_r U(x,\mu)_{t,s,r}U(x,\mu)_{t,s,r}^{-1}$ in
eq.~(\ref{twoxtwentyfive}) and then take a derivative of
eq.~(\ref{twoxtwentyfive}) with respect to the parameter~$s$. By
noting the property $\Tr(\delta_1\hat P_-\delta_2\hat P_-\delta_3\hat
P_-)=0$,\footnote{This can be proven by inserting $1=(\gamma_5)^2$ and
  using $\{\hat\gamma_5,\delta\hat P_-\}=0$.} one has
\begin{equation}
   \partial_t\left[\partial_s\mathfrak{L}_\zeta
     -i\Tr(P_{t,s,r}[\partial_s P_{t,s,r},\partial_rP_{t,s,r}])\right]
   -\partial_r\left[\partial_s\mathfrak{L}_\eta
     -i\Tr(P_{t,s,r}[\partial_sP_{t,s,r},\partial_tP_{t,s,r}])\right]=0\,,
\label{twoxfortyseven}
\end{equation}
where $P_{t,s,r}=\hat P_-|_{U=U_{t,s,r}}$. A particular solution of
this equation is given by $\partial_s\mathfrak{L}_\eta= i\Tr(\hat
P_{t,s,r}[\partial_s\hat P_{t,s,r},\partial_t\hat P_{t,s,r}])$. Then
by getting rid of the parameter~$r$ and integrating this relation
along a direction of~$s$ in figure~\ref{fig3}, we have
\begin{equation}
   \mathfrak{L}_\eta= i\int_0^1\rmd
   s\Tr(P_{t,s}[\partial_sP_{t,s},\delta_\eta P_{t,s}])\,,
\label{twoxfortyeight}
\end{equation}
where $P_{t,s}=\hat P_-|_{U=U_{t,s}}$. Conversely, it is
straightforward to verify that this measure term in fact satisfies the
local integrability~(\ref{twoxtwentyfive}). This form of the measure
term already appeared in
refs.~\cite{Luscher:1999du,Suzuki:1999qw}. This measure term is given
by a combination of the Dirac operator and the corresponding measure
current is a local function of link variables due to the locality of
the Dirac operator. Thus it defines a sensible chiral determinant at
least for configurations restricted as above.

The measure term~(\ref{twoxfortyeight}) is yet ambiguous depending on
the choice of the two-pa\-ra\-me\-ter family~$U_{t,s}$. If we consider an
infinitesimal deformation of the two-parameter family~$\delta
U_{t,s}$, we find the the measure term changes by
\begin{equation}
   \delta\mathfrak{L}_\eta= i\delta_\eta\int_0^1\rmd
   s\Tr(P_{t,s}[\partial_sP_{t,s},\delta P_{t,s}])\,.
\label{twoxfortynine}
\end{equation}
By comparing this with the variation of the chiral
determinant~(\ref{twoxtwentysix}), we realize that this ambiguity is
nothing but the standard ambiguity in defining the effective action, a
freedom of choosing local counterterms (coboundary terms). To avoid
this unimportant ambiguity, it is convenient to fix a convention for
the two-parameter family~$U_{t,s}$. Here we take
eq.~(\ref{twoxfortysix}) as a part of definition of the WZW term. By
this way, we have
\begin{equation}
   \Gammait_{\rm WZW}[g,U] =a^{2n}\sum_x\mathcal{A}^a(x)|_U
   \int_0^1\rmd t\,(\partial_t g_tg_t^{-1})^a(x) -i\int_0^1\rmd
   t\int_0^1\rmd s \Tr(P_{t,s}[\partial_tP_{t,s},\partial_sP_{t,s}])\,,
\label{twoxfifty}
\end{equation}
where $P_{t,s}=\hat P_-|_{U=U_{t,s}}$ and $U_{t,s}$ is defined by
eq.~(\ref{twoxfortysix}).  This completes our construction of the WZW
term on the lattice.\footnote{Once the WZW term in this form has been
  obtained, we may get rid of the chiral
  determinant~(\ref{twoxone}). In particular, this form works even
  when the Dirac operator has accidental and/or topological zero
  modes. The WZW term is defined also for topologically non-trivial
  sectors, if we change eq.~(\ref{twoxfortysix}) appropriately, as
  $U_{t,s}(x,\mu)=
  U_0(x,\mu)^{(1-s)}[g_t^{-1}(x)U(x,\mu)g_t(x+a\hat\mu)]^s$, where
  $U_0(x,\mu)$ is a reference configuration in the topological
  sector. A more satisfactory derivation of eq.~(\ref{twoxfifty}) is
  obtained by starting with the expectation
  value~$\langle\mathcal{O}\rangle_{\rm F}$, instead of the chiral
  determinant.} In appendix~\ref{section8}, we show that this
definition has the correct classical continuum limit.

The expression~(\ref{twoxfifty}) can be cast into a parametrizaion
free form as
\begin{equation}
   \Gammait_{\rm WZW}[g,U] =-i\int_C\Tr PD^{-1}P_+\rmd
   D-i\int_{D^2}\Tr P\rmd P\rmd P\,,
\label{twoxfiftyone}
\end{equation}
where $D^2$ is the 2-disk $D^2$ in the space of admissible gauge
fields parametrized by $t$ and~$s$ as depicted in figure~\ref{fig3}, $C$ is the
path connecting $U$ and $U^{g^{-1}}$ parametrized by $t$ and $\rmd$ is
the exterior derivative.  The Dirac operator and the projector on the
disk $D^2$ are simply denoted by $D$ and $P$. Since $C$ is a path
along the gauge direction, the first term on the right hand side
of~(\ref{twoxfiftyone}) is local in the sense of lattice theory. As we
shall show in the next section, the WZW term modulo $2\pi$ does not
depend on the choice of $D^2$ as far as the two boundaries
parametrized by~$s$ are fixed and $C$ lies in the gauge direction.

By using eq.~(\ref{twoxfifty}) and the
relation~$U^{(hg)^{-1}}=(U^{h^{-1}})^{g^{-1}}$, we find the
composition law:
\begin{equation}
   \Gammait_{\rm WZW}[hg,U] =\Gammait_{\rm WZW}[h,U]+\Gammait_{\rm
     WZW}[g,U^{h^{-1}}]\pmod{2\pi}\,,
\label{twoxfiftytwo}
\end{equation}
where the ambiguity of~$2\pi$ will be explained in the next subsection
(each WZW term in the above expression has such an ambiguity).

\section{$2\pi$-ambiguity of the WZW term and the Witten term}\label{section3}

\EPSFIGURE[t]{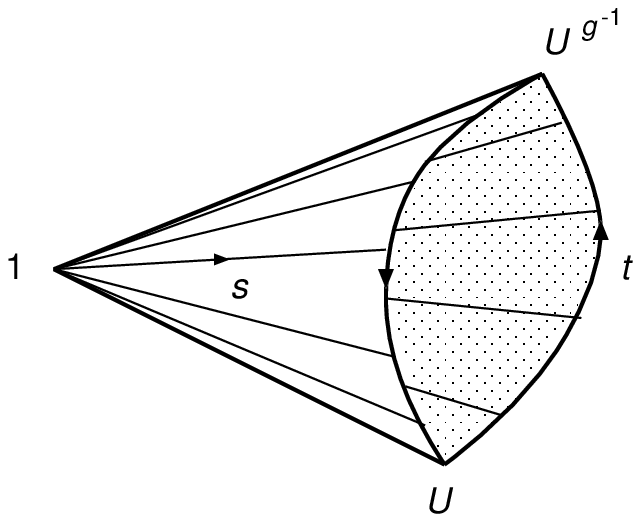,width=6.5cm}{A difference between WZW terms defined with
respect to two different gauge paths.\label{fig4}}

The WZW term~(\ref{twoxfifty}) depends on a chosen gauge path~$g_t$
through eq.~(\ref{twoxfortysix}) and this freedom of gauge path causes
a $2\pi$-ambiguity of the WZW term. To show this, let us consider two
gauge paths, $g_t^{(1)}$ and~$g_t^{(2)}$. The difference between the
WZW terms defined with respect to each path is obviously given by
$\Gammait_{\rm WZW}[1,U]$ in which the gauge path~$g_t$ ($g_0=g_1=1$)
is defined by\footnote{Note that the WZW term is $t$-reparametrization
  invariant.}
\begin{equation}
   g_t(x)=\cases{g_{2t}^{(1)}(x)\,,&for $0\leq t\leq\displaystyle\frac{1}{2}$\,,\cr\cr
                 g_{2(1-t)}^{(2)}(x)\,,&for $\displaystyle\frac{1}{2}\leq t\leq1$\cr}
\label{threexone}
\end{equation}
which forms a closed loop. See figure~\ref{fig4}. However, we have
observed in eq.~(\ref{twoxfortytwo}) that, along a gauge
loop, the line integral of the covariant anomaly in the WZW term is
given by the surface integral of the curvature over a surface spanned
by the loop. As a consequence, the difference is given by
\begin{equation}
   \Gammait_{\rm WZW}[g,U]|_{g_t^{(1)}}-\Gammait_{\rm
     WZW}[g,U]|_{g_t^{(2)}} =-\int_{S^2}\rmd t\,\rmd s\,
   i\Tr(P_{t,s}[\partial_tP_{t,s},\partial_sP_{t,s}])\,,
\label{threextwo}
\end{equation}
here the integral is for a surface of the two dimensional cone
depicted in figure~\ref{fig4}, which is topologically a
2-sphere~$S^2$. Note that an orientation of the coordinate
system~$(t,s)$ in figure~\ref{fig2} is opposite compared to that in
figure~\ref{fig4}.  Here we adopt a convention that $(t,s)$ is the right-handed
system. Thus the difference coincides with ($-2\pi$ times) the first
Chern number~(\ref{twoxtwentytwo}) for the surface of the cone,
$\mathcal{M}=S^2$.  Therefore, even if the difference exists, it is a
multiple of~$2\pi$ and $\exp(i\Gammait_{\rm WZW}[g,U])$ is a single
valued functional of $g$ and~$U$. If the Chern number is non-zero, the
2-sphere is non-contractible in the space of admissible gauge
fields~$\mathfrak{U}$. An important implication of this fact will be
discussed in the next section.

Next we write the WZW term as
\begin{equation}
   \Gammait_{\rm WZW}[g,U]=\Gammait_{\rm WZW}[g,U]-\Gammait_{\rm
     WZW}[g,1] +\Gammait_{\rm W}[g]\,,
\label{threexthree}
\end{equation}
where $\Gammait_{\rm W}[g]=\Gammait_{\rm WZW}[g,1]$ will be referred
to as the Witten term. The following simple argument shows that the
part $\Gammait_{\rm WZW}[g,U]-\Gammait_{\rm WZW}[g,1]$ is free of the
above $2\pi$-ambiguity. Namely it depends only on $g$ and~$U$ and is
ignorant about the ``history''~$g_t$ how $g$ evolved from~$1$. The
lattice Witten term~$\Gammait_{\rm W}[g]$, on the other hand, carries
the $2\pi$-ambiguity and depends only on the gauge
transformations~$g_t$. To see these facts, let us consider how the WZW
term $\Gammait_{\rm WZW}[1,U]$ for a gauge loop~$g_t$ ($g_0=g_1=1$)
changes as we deform the gauge field~$U$ to the trivial one,
$U\to1$. Defining a one-parameter family of gauge fields~$U_r$ such
that $U_0=U$ and~$U_1=1$,\footnote{For example,
  $U_r(x,\mu)=[U(x,\mu)]^{1-r}$ is enough.} at each value of the
parameter~$r$, we have a cone depicted figure~\ref{fig4} in which $U$
is replaced by~$U_r$. The point is that everywhere of the
\emph{surface\/} of this cone is admissible by our assumption on $U$
and~$g$ and, therefore, in $\Gammait_{\rm WZW}[1,U]$ we can
deform~$U\to1$ without encountering any violation of the
admissibility. Since the Chern number, being an integer, is invariant
under such a smooth deformation, the Chern numbers are identical for
the two closed surfaces in figure~\ref{fig5}. This implies that
$\Gammait_{\rm WZW}[g,U]$ and~$\Gammait_{\rm WZW}[g,1]$ has an
identical $2\pi$-ambiguity; if one changes the path~$g_t$,
$\Gammait_{\rm WZW}[g,U]$ and~$\Gammait_{\rm WZW}[g,1]$ exhibit an
identical shift of~$2\pi$. This shows the above assertions.

\FIGURE[t]{\centerline{\epsfig{file=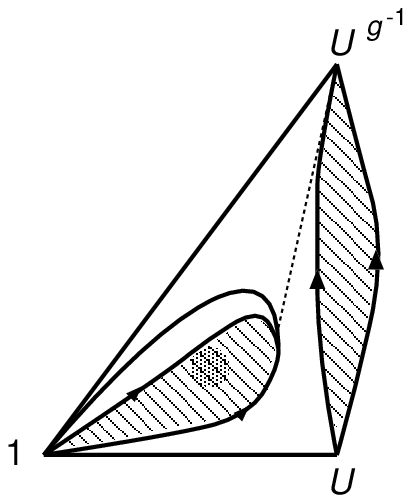,width=5.5cm,clip=}}%
\caption{An integration of the curvature over the larger and the
  smaller closed surfaces gives rise to an integer multiple of $2\pi$
  associated to an ambiguity of~$\Gammait_{\rm WZW}[g,U]$ and
  of~$\Gammait_{\rm WZW}[g,1]$ respectively. These two surfaces can be
  deformed to each other without encountering any violation of the
  admissibility. The shaded small circle contained in the smaller
  surface indicates a possible violation of the
  admissibility.\label{fig5}}}

In the continuum theory, the WZW term is naturally decomposed into the
$2n$ dimensional part, which depends only on the boundary values, $g$
and~$U$, and the $2n+1$ dimensional Witten term, which does depend
on~$g_t$ but not on the gauge field, through the ``homotopy
formula''~(\ref{axeighteen}). It is interesting that our argument
above performs this decomposition in effect, although we do not know a
lattice analogue of the homotopy formula for the present.

\section{Gauge anomaly and an obstruction for a smooth measure over\label{section4}
$\mathfrak{U}$}

Since information concerning the gauge anomaly is gathered in the WZW
term, it is natural to examine the winding number of the phase of
chiral determinant along a gauge loop~\cite{Alvarez-Gaume:1983cs} in
our formulation. The winding number is obtained by considering a gauge
loop~$g_t$ such that $g_0=g_1=1$.  Repeating the argument in the
previous section, we have as the winding number
\begin{equation}
   {1\over2\pi}\Gammait_{\rm WZW}[1,U] =-{1\over2\pi}\int_{S^2}\rmd
   t\,\rmd s\, i\Tr(P_{t,s}[\partial_tP_{t,s},\partial_sP_{t,s}])\,,
\label{fourxone}
\end{equation}
where the integral is over a surface of the two dimensional cone (or
$S^2$) depicted in figure~\ref{fig4}. The above expression is nothing
but the first Chern number~(\ref{twoxtwentytwo}) for $\mathcal{M}=S^2$
which is an integer. To see when this number becomes non-zero, it is
enough to see the classical continuum limit, because it is an integer
which cannot depend on the lattice spacing~$a$.  {}From
eq.~(\ref{axtwenty}) in appendix~\ref{section8}, we have
\begin{eqnarray}
   {1\over2\pi}\Gammait_{\rm WZW}[1,U]
   &=&-{(-1)^ni^{n+1}n!\over(2\pi)^{n+1}(2n+1)!}
   \int_{S^{2n}\times S^1}\rmd^{2n+1}x\times
\nonumber\\&&
   \times\, \epsilon_{\mu_1\mu_2\cdots\mu_{2n+1}}
   \tr[R(g_t^{-1}\partial_{\mu_1}g_t)R(g_t^{-1}\partial_{\mu_2}g_t)\cdots
     R(g_t^{-1}\partial_{\mu_{2n+1}}g_t)]\,.
\label{fourxtwo}
\end{eqnarray}
The integrand is a total divergence and thus the integral is a
homotopy invariant of the mapping~$g_t(x)$ from~$S^{2n}\times S^1$ to
the gauge group~$G$. However, since $G$ is simply-connected, the
homotopy class is identical to that of mappings from the
sphere~$S^{2n+1}$ to~$G$~\cite{Alvarez-Gaume:1983cs}. Therefore, when
the mapping~$g_t(x)$ corresponds to a non-trivial element
of~$\pi_{2n+1}(G)$, the above winding number may become non-zero,
depending on the representation~$R$. A non-trivial $\pi_{2n+1}(G)$ is
a measure for the gauge anomaly in the continuum
theory~\cite{Alvarez-Gaume:1983cs}. These are basically observations
already made in refs.~\cite{Adams:1999um,Adams:2000yi}.

\looseness=1 Our argument, however, shows that the gauge anomaly has a profound
implication in the present lattice formulation. Eq.~(\ref{fourxone})
says that when the winding number, which is a measure of the gauge
anomaly, is non-zero, the Chern number~(\ref{twoxtwentytwo}) defined
for a surface of the cone in figure~\ref{fig4} becomes non-zero. This implies
that the $\U(1)$ bundle associated to the fermion measure is
non-trivial and, as we discussed in section~\ref{section2.2}, and
\emph{the Weyl fermion cannot consistently be formulated, even if one
  sacrifices the gauge invariance.} \pagebreak[3] This shows that anomalous cases
are completely different from anomaly-free cases from a view point of
the present lattice formulation based on the GW relation. It seems
interesting to re-examine quantization of anomalous gauge
theories~\cite{Faddeev:pc} in light of this observation.

\section{Witten term and the multivaluedness}\label{section5}

\EPSFIGURE[t]{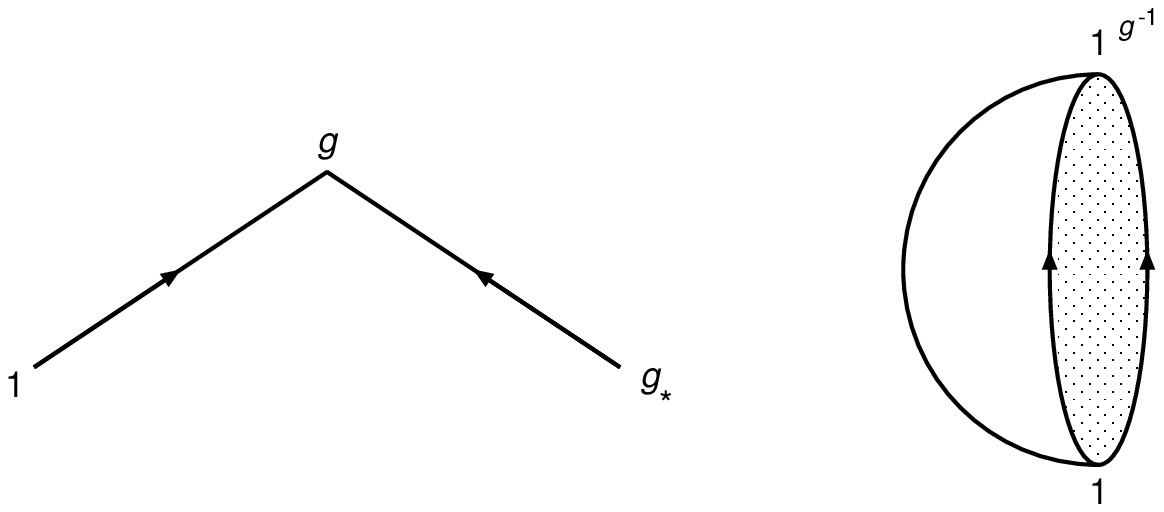}{The two different paths to define the Witten term
  in the space of gauge transformations (left) and in the space of
  gauge field configurations (right). A difference between the Witten
  terms is given by an integral of the curvature over the closed
  surface in the right-hand figure.\label{fig6}}

We defined the Witten term by
\begin{equation}
   \Gammait_{\rm W}[g]=\Gammait_{\rm WZW}[g,1]\,.
\label{fivexone}
\end{equation}
According to an understanding in the continuum
theory~\cite{Witten:tw}, this term should have an ambiguity of~$2\pi$
(multivaluedness) in a somewhat different sense from what we discussed
so far. The key observation is that when $U=1$, the choice of the
gauge path~(\ref{twoxtwentyseven}) has a wider freedom. Let $g_*$ be a
\emph{constant\/} $G$-transformation. Since $1^{g_*}=1$, one can use
$g_t(x)$ which has the boundary values
\begin{equation}
   g_0(x)=g_*,\qquad g_1(x)=g(x)\,,
\label{fivextwo}
\end{equation}
instead of eq.~(\ref{twoxtwentyeight}) in defining of~$\Gammait_{\rm
  W}[g]$.  Then in the space of gauge transformations, we consider two
paths, one starts from~$1$ and ends at~$g$ and another starts
from~$g_*$ and ends with~$g$. See figure~\ref{fig6}. We can then consider a
difference between the Witten terms defined by each path. This is
given by the Witten term defined by a path in the space of gauge
transformations, which starts at $1$ and ends at~$g_*$. In the space
of gauge fields, this path corresponds to a \emph{loop}. Even with the
above change~(\ref{fivextwo}), our arguments so far hold with trivial
changes. For example, $1$ in the right hand bottom corner of the
square in figure~\ref{fig2} is replaced by~$g_*$. The
relation~(\ref{twoxfortytwo}) for a loop holds as it stands and, by a
similar argument as above, we have
\begin{equation}
   \Gammait_{\rm W}[g]|_1-\Gammait_{\rm W}[g]|_{g_*}
   =-\int_{S^2}\rmd t\,\rmd s\,
   i\Tr(P_{t,s}[\partial_tP_{t,s},\partial_sP_{t,s}])\,,
\label{fivexthree}
\end{equation}
where we have indicated the initial value in two gauge paths. This is
an integer multiple of~$2\pi$ because it is given by the Chern
number~(\ref{twoxtwentytwo}). The explicit value of this difference
can be found by the classical continuum limit
\begin{eqnarray}
   \Gammait_{\rm W}[g]|_1-\Gammait_{\rm W}[g]|_{g_*}
   &=&-{(-1)^ni^{n+1}n!\over(2\pi)^n(2n+1)!}
   \int_{S^{2n}\times I}\rmd^{2n+1}x\times
\nonumber\\&&
\times\, \epsilon_{\mu_1\mu_2\cdots\mu_{2n+1}}
\tr[R(g_t^{-1}\partial_{\mu_1}g_t)R(g_t^{-1}\partial_{\mu_2}g_t)\cdots
  R(g_t^{-1}\partial_{\mu_{2n+1}}g_t)]\,,\qquad
\label{fivexfour}
\end{eqnarray}
where the boundary condition along the interval~$I$ is such that
$g_0=1$ and~$g_1=g_*$. Then the picture in figure~\ref{fig7}
emerges. Namely, the difference is given by ($-2\pi$ times) a winding
number of the mapping~$g_t(x)$ which wraps around a basic $2n+1$
dimensional basic sphere in~$G$~\cite{Witten:tw}.\footnote{To be
  precise, this statement is true when the representation~$R$ is the
  defining representation of~$G$. Depending on the representation~$R$,
  the right hand side of eq.~(\ref{fivexfour}) may be a multiple of
  the winding number.} By this way, a non-trivial~$\pi_{2n+1}(G)$
implies the multivaluedness of the Witten term on the lattice by
$2\pi$, as in the continuum~\cite{Witten:tw}.

\EPSFIGURE[t]{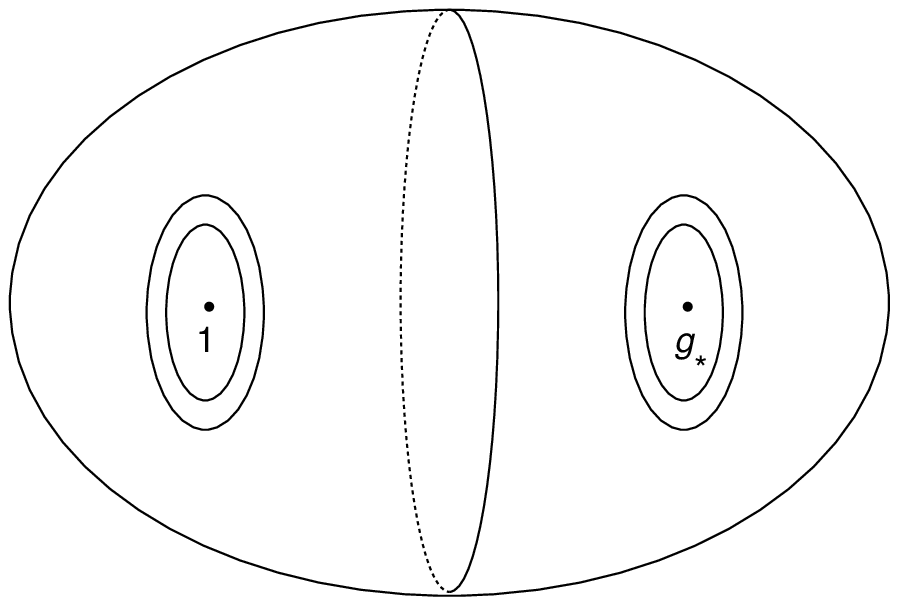,width=.6\textwidth}{The image of the mapping~$g_t(x)$ from $S^{2n}\times I$
to~$G$ in eq.~(\ref{fivexfour}). Each latitude corresponds to the image of
$g_t(x)$ at each~$t$. The integral~(\ref{fivexfour}) is the winding number on
this $2n+1$~dimensional sphere in~$G$.\label{fig7}}

\section{Global $\SU(2)$ anomaly}\label{section6}

As the final application of the WZW term, we consider the global
$\SU(2)$ anomaly~\cite{Witten:fp} in lattice gauge
theory~\cite{Neuberger:1998wg,Bar:1999ka}. We assume that $d=2n=4$ and
the Weyl fermion belongs to the fundamental ${1\over2}$ representation
of~$\SU(2)$.  Since $\pi_4(\SU(2))=Z_2$, there exists one homotopy
class of gauge transformations in the continuum theory which cannot
smoothly be deformed into the identity. Let $g(x)$ be such a
homotopically non-trivial gauge transformation. In the continuum
theory, Witten showed that the chiral determinant for $A^g$, the gauge
transformation of the vector potential~$A$, has a different sign
compared to that for~$A$. This implies $\Gammait_{\rm
  WZW}[g,U]=\pi\bmod 2\pi$ in terms of the WZW term.

In the present case, since the fundamental representation of~$\SU(2)$
is pseudo-real, we have\footnote{To show these relations, one has to
  assume the charge conjugation property of the Dirac operator,
  $D(x,y)^T=CD_{R\to R^*}(x,y)C^{-1}$, where $R^*$ is the complex
  conjugate representation of~$R$ and $C$ is the charge conjugation
  matrix. The overlap-Dirac operator possesses this property. The
  fundamental representation of $\SU(2)$ is pseudo-real, i.e., in the
  standard convention, one has
  $R(T^a)^*=-R(T^a)^T=\tau_2R(T^a)\tau_2$. So by defining
  $B=C\gamma_5\tau_2$, one has $BR(T^a)B^{-1}=-R(T^a)^T$,
  $B\gamma_5B^{-1}=\gamma_5^T$ and
  $B\gamma_5D(x,y)B^{-1}=[\gamma_5D(x,y)]^T$. Using these properties,
  it is straightforward to show these relations.}
\begin{equation}
   \mathcal{A}^a(x)=0\,,\qquad
   i\Tr(\hat P_-[\delta_\eta\hat P_-,\delta_\zeta\hat P_-])=0\,.
\label{sixxone}
\end{equation}
So \emph{if\/} our construction~(\ref{twoxfifty}) is applicable to
this case, we would conclude~$\Gammait_{\rm WZW}[g,U]=0$ which does
not reproduce the $\SU(2)$ anomaly. In fact, our construction breaks
down due to a non-contractible \emph{loop\/} in the space of
admissible gauge fields~$\mathfrak{U}$~\cite{Bar:1999ka}. The point is
that each configuration on intermediate points of the gauge
path~(\ref{twoxtwentyseven}), where $g$ is given by a lattice
restriction of the continuum~$g$, does not have a continuum
counterpart, because~$g$ in the continuum is not homotopically
equivalent to the identity. This implies that there is no guarantee
that the whole disk in figure~\ref{fig3} is contained in~$\mathfrak{U}$. In fact,
the analysis of ref.~\cite{Bar:1999ka} shows that there must be a
``hole'' on the disk. The contribution of this hole must be taken into
account to pick up the correct phase in the WZW term (a situation
analogous to the Aharanov-Bohm effect).

This difficulty can be evaded by enlarging the gauge group~$\SU(2)$ to
a larger group~$G$ for which $\pi_4(G)=0$, say $G=\SU(3)$. We then
consider the fundamental representation of $\SU(3)$ and embed the
fundamental representation of~$\SU(2)$
as~\cite{Witten:tw,Elitzur:1984kr,Bar:1999ka}
\begin{equation}
   \Omega(x)=\pmatrix{g(x)&0\cr 0&1\cr}\in\SU(3).
\label{sixxtwo}
\end{equation}
Since $\pi_4(\SU(3))=0$, there exists a one-parameter family of
$\SU(3)$ gauge transformations in the continuum theory which
interpolates $1$ and $\Omega(x)$.  The lattice transcription of these
$\SU(3)$ gauge transformations thus provides a gauge
path~$\Omega_t(x)$ such that $\Omega_0(x)=1$
and~$\Omega_1(x)=\Omega(x)$ and each point of~$\Omega_t(x)$ has the
continuum counterpart. Therefore our construction of the WZW term can
be applied along this gauge path~$\Omega_t(x)$ in $\SU(3)$ lattice
theory. See figure~\ref{fig8}.

\FIGURE[t]{\centerline{\epsfig{file=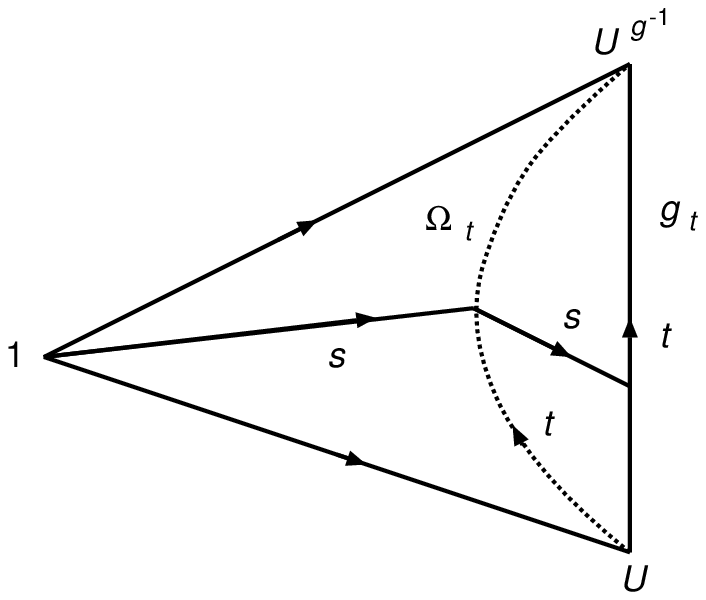,width=8cm}}%
\caption{A way to construct the WZW term corresponding to a
  fundamental fermion of~$\SU(2)$. We can apply our construction of
  the measure term and the WZW term along the gauge path of~$\SU(3)$,
  $\Omega_t$ (the broken line). The measure term for the $\SU(2)$
  gauge path~$g_t$ (the straight line) is defined by extending the
  $s$-integration along the gauge surface spanned by two gauge
  paths. Then the WZW term defined with respect to the $\SU(2)$ gauge
  path is obtained. These two WZW terms, however, are
  identical.\label{fig8}}}

In the $\SU(3)$ theory, we then have two gauge paths, one is nothing
but $g_t$ trivially embedded in the $\SU(3)$ theory and another is
given by $\Omega_t$.  As shown in figure~\ref{fig8}, within the
$\SU(3)$ theory, we can provide measure terms for those two gauge
paths and define the WZW term along each of them. However, by
repeating an argument similar to that of section~\ref{section3}, we
can see that a difference between those two WZW terms is zero. Thus we
define the WZW term corresponding to a fundamental fermion
of~$\SU(2)$, $\Gammait_{\rm WZW}[g,U]$ by using eqs.~(\ref{twoxfifty})
and~(\ref{twoxfortysix}) with the replacement~$g_t\to\Omega_t$.

We next show that
\begin{equation}
   \Gammait_{\rm WZW}[g,U]=\Gammait_{\rm WZW}[g,1]=\Gammait_{\rm
     W}[g]\,.
\label{sixxthree}
\end{equation}
To show this, we introduce a one-parameter family of $\SU(2)$ gauge
fields~$U_r$ such that $U_0=U$ and~$U_1=1$ and consider $\Gammait_{\rm
  WZW}[g,U_r]$. According to our construction of the WZW term,
$\Gammait_{\rm WZW}[g,U_r]$ is given by an integral of the curvature
over a surface~$\mathcal{S}_r$ whose vertices are $1$, $U_r$
and~$U_r^{g^{-1}}$, as depicted in figure~\ref{fig8} for~$r=0$. We note that all
configurations on these surfaces~$\mathcal{S}_r$ are admissible for
every~$r$. We can therefore deform the surface~$\mathcal{S}_0$, to
that for~$r=1$, $\mathcal{S}_1$, without encountering any violation of
the admissibility as depicted in figure~\ref{fig9}. These facts show that a
difference between the WZW terms, $\Gammait_{\rm WZW}[g,U]$
and~$\Gammait_{\rm WZW}[g,1]$, is given by an integral of the
curvature over a surface~$\mathcal{R}$ which is swept by the boundary
of~$\mathcal{S}_r$ in the process of the above deformation (the shaded
area of figure~\ref{fig9}), because $\mathcal{S}_0$, $\mathcal{S}_1$ and
$\mathcal{R}$ form a closed 2-surface and within the closed surface
there is no violation of the admissibility. However, the integral of
the curvature over $\mathcal{R}$ vanishes because configurations
on~$\mathcal{R}$ act on the fundamental representation of $\SU(3)$ as
${1\over2}\otimes 1$ representation of~$\SU(2)$ and for such a
representation, the curvature vanishes as eq.~(\ref{sixxone})
shows. This argument establishes eq.~(\ref{sixxthree}).

\EPSFIGURE{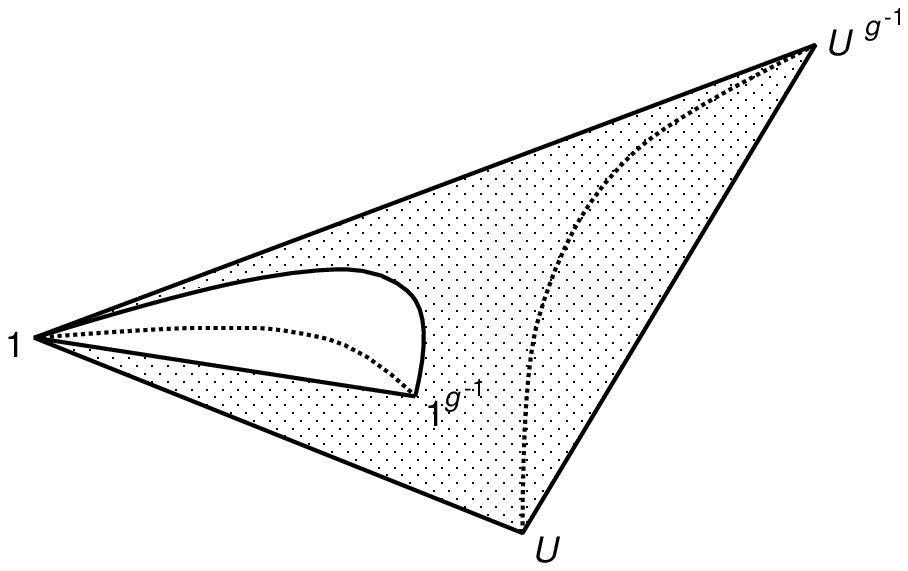}{Surfaces~$\mathcal{S}_0$ and $\mathcal{S}_1$, and
the region~$\mathcal{R}$ (the shaded area) which is swept by the boundary
of~$\mathcal{S}_r$, utilized to show eq.~(\ref{sixxthree}).\label{fig9}}

By a similar deformation argument, one can see that
\begin{equation}
   \Gammait_{\rm WZW}[g,U]=\Gammait_{\rm WZW}[h,U],
\label{sixxfour}
\end{equation}
as long as $g$ and $h$ are given by a lattice restriction of
homotopically equivalent $\SU(2)$ gauge transformations~$g(x)$
and~$h(x)$ in the continuum.

The above two properties of the WZW term, combined with the classical
continuum limit, show that $\Gammait_{\rm WZW}[g,U]=\pi\bmod 2\pi$. We
first note that $g(x)^2$ in the continuum belongs to a trivial
homotopy class and according to eqs.~(\ref{sixxthree})
and~(\ref{sixxfour}),
\begin{equation}
   \Gammait_{\rm WZW}[g^2,U]=\Gammait_{\rm WZW}[1,U] =\Gammait_{\rm
     W}[1]=0\pmod{2\pi}\,,
\label{sixxfive}
\end{equation}
where the last equality follows from the argument of
section~\ref{section3}. On the other hand, from the composition
law~(\ref{twoxfiftytwo}) and eq.~(\ref{sixxthree}), we have
\begin{eqnarray}
   \Gammait_{\rm WZW}[g^2,U]
   &=&\Gammait_{\rm WZW}[g,U]+\Gammait_{\rm WZW}[g,U^{g^{-1}}]\pmod{2\pi}
\nonumber\\
   &=&2\Gammait_{\rm W}[g]\pmod{2\pi}\,.
\label{sixxsix}
\end{eqnarray}
Then from eqs.~(\ref{sixxfive}) and~(\ref{sixxsix}), we infer that
possible values of the WZW term are \emph{quantized\/} even on the
lattice as $\Gammait_{\rm WZW}[g,U]=\Gammait_{\rm W}[g]=0\bmod2\pi$ or
$\Gammait_{\rm WZW}[g,U]=\Gammait_{\rm W}[g]=\pi\bmod2\pi$. One can
decide which is actually the case from the classical continuum
limit~(\ref{axtwentyone})
\begin{equation}
   \Gammait_{\rm W}[g] ={i\over240\pi^2}\int_{S^4\times I}\rmd^5 x\,
   \epsilon_{\mu\nu\rho\sigma\lambda}
   \tr[R(\Omega_t^{-1}\partial_\mu\Omega_t)
     R(\Omega_t^{-1}\partial_\nu\Omega_t)\cdots
     R(\Omega_t^{-1}\partial_\lambda\Omega_t)]+O(a)\,,
\label{sixxseven}
\end{equation}
where $R$ stands for the fundamental representation
of~$\SU(3)$. According to~ref.~\cite{Witten:tx}, the above integral
is~$\pi\bmod2\pi$ and thus $\Gammait_{\rm WZW}[g,U]=\pi\bmod2\pi$. By
this way, the lattice Witten term reproduces the $\SU(2)$ anomaly.

\section{Conclusion}\label{section7}

In this paper, we have formulated the WZW term in lattice gauge theory
on the basis of a Dirac operator which satisfies the GW relation. We
observed that topological properties of the WZW term in the continuum
are neatly reproduced in the lattice framework, although a basic
mechanism is rather different: In our formulation, a non-trivial
topology of (a connected component of) the \emph{configuration
  space\/} of lattice gauge fields due to the admissibility is the key
element. On the other hand, (a connected component of) the
configuration space of gauge fields in the continuum is topologically
trivial.  An important observation which emerged through our
analysis\footnote{See also ref.~\cite{Adams:2002tb}.} is that it is
generically impossible to formulate a Weyl fermion in an anomalous
gauge-group representation, along the line of
refs.~\cite{Luscher:1999du,Luscher:1999un}, because for such a
representation, the Chern number~(\ref{fourxone}) may become
non-zero. If this occurs, a smooth fermion measure does not exist even
the gauge invariance is sacrificed. With this observation, it seems
interesting to re-examine a quantization of anomalous chiral gauge
theories~\cite{Faddeev:pc} within the lattice framework on the basis
of the GW relation.

\acknowledgments

We would like to thank Hideaki Ohshima for explaining us on the
homotopy class of mappings from torus. H.S.\ would like to thank
Oliver B\"ar and David H.~Adams for useful discussions. This work is
supported in part by Grant-in-Aid for Scientific Research, \#13135203
and~\#13640258.

\appendix

\section[Classical continuum limit]{Classical continuum
  limit\protect\footnote{In this appendix, we will
denote $R(T^a)$ by~$T^a$ for notational simplicity.}}
\label{section8}

The lattice transcription~(\ref{twoxfortyfive}) of a continuum gauge
field is assumed in taking the classical continuum limit. To write
down various quantities in the continuum limit, a use of differential
form is quite helpful.  The gauge potential 1-form and the field
strength 2-form are defined respectively by
\begin{equation}
   A=A_\mu\rmd x_\mu\,,\qquad F=\rmd A+A^2={1\over2}F_{\mu\nu}\rmd
   x_\mu\rmd x_\nu
\label{axone}
\end{equation}
from a gauge field in the continuum theory. The exterior
derivative~$\rmd$ is defined by~$\rmd=\rmd x_\mu\,\partial/(\partial
x_\mu)$.

It is not so difficult to explicitly evaluate the continuum limit of
the covariant ano\-maly~(\ref{twoxthirtyfive}) for the overlap-Dirac
operator~\cite{Kikukawa:1998pd}. For an evaluation in arbitrary even
dimensions, see ref.~\cite{Fujiwara:2002xh}. By multiplying the volume
form~$\rmd^{2n}x=\rmd x_1\cdots\rmd x_{2n}$, the continuum limit is
given by\footnote{Our gamma matrices are hermitean,
  $\gamma_\mu^\dagger=\gamma_\mu$ and
  $\{\gamma_\mu,\gamma_\nu\}=2\delta_{\mu\nu}$. The chiral
  matrix~$\gamma_5$ is defined by
  $\gamma_5=(-i)^n\gamma_1\cdots\gamma_{2n}$ and
  $\gamma_5^\dagger=\gamma_5$ and $\gamma_5^2=1$.}
\begin{equation}
   \mathcal{A}^a(x)\,\rmd^{2n}x=c_1\tr[T^aF(x)^n]+O(a)\,,
\label{axtwo}
\end{equation}
where
\begin{equation}
   c_1=-{i^{n+1}\over(2\pi)^nn!}\,.
\label{axthree}
\end{equation}
On the other hand, an explicit evaluation of the continuum limit of
the curvature \linebreak $i\Tr(\hat P_-[\delta_\eta\hat
  P_-,\delta_\zeta\hat P_-])$ is more involved, although
executable~\cite{Adams:2000yi}. Here we invoke a general
argument~\cite{Luscher:1999un} which immediately leads to the
answer. If the variations $\eta_\mu(x)$ and~$\zeta_\mu(x)$ are lattice
restrictions of some smooth vector fields in the continuum, the
locality and symmetries of the curvature imply that a general form of
the continuum limit is given by
\begin{equation}
   i\Tr(\hat P_-[\delta_\eta\hat P_-,\delta_\zeta\hat P_-])
   =c_2\int_{T^{2n}}\str(\eta\zeta F^{n-1})+O(a)\,,
\label{axfour}
\end{equation}
where $\eta=\eta_\mu\rmd x_\mu$, $\zeta=\zeta_\mu\rmd x_\mu$ and
$\str$ denotes the symmetrized trace~\cite{Zumino:1983rz} defined by
\begin{equation}
   \str(A_1A_2\cdots A_n) =\sum_\sigma{1\over n!}\epsilon_\sigma
   \tr(A_{\sigma(1)}A_{\sigma(2)}\cdots A_{\sigma(n)})\,.
\label{axfive}
\end{equation}
In this expression, the summation is taken over all
permutations~$\sigma$ and $\epsilon_\sigma$ is the parity associated
to the permutation~$\sigma$, $A_1A_2\cdots A_n\to
A_{\sigma(1)}A_{\sigma(2)}\cdots A_{\sigma(n)}$, regarding odd rank
forms as an anti-commuting quantity. Then by putting
eqs.~(\ref{axtwo}) and~(\ref{axfour}) into eq.~(\ref{twoxthirtysix}),
where $\eta=-\rmd\omega-[A,\omega]+O(a)$ and $\delta_\zeta
A=\zeta+O(a)$ in the continuum limit, we find
\begin{equation}
   c_2=-nc_1
\label{axsix}
\end{equation}
by using the Bianchi identity
\begin{equation}
   \rmd F+[A,F]=0\,.
\label{axseven}
\end{equation}

In the continuum limit, the two parameter family of the gauge
fields~(\ref{twoxfortysix}) corresponds to a family of the gauge
potentials
\begin{equation}
   A_{t,s}=sA^{g_t^{-1}}+O(a),\qquad A^{g_t^{-1}}=g_t^{-1}(\rmd+A)g_t\,.
\label{axeight}
\end{equation}
Then in eq.~(\ref{twoxfifty}), a contribution of the covariant anomaly
to the WZW term $\Gammait_{\rm WZW}$ becomes
\begin{equation}
   c_1\int_{T^{2n}}\int_0^1\rmd t\, \tr(g_t^{-1}\partial_t
   g_tF^n)|_{A=A^{g_t^{-1}}}+O(a)\,.
\label{axnine}
\end{equation}
On the other hand, the contribution of the measure term is given by
substituting
\begin{eqnarray}
   \eta&=&\partial_t A_{t,s}
   =s[\rmd(g_t^{-1}\partial_t g_t)+[A^{g_t^{-1}},g_t^{-1}\partial_t g_t]],
\nonumber\\
   \zeta&=&\partial_s A_{t,s}=A^{g_t^{-1}},
\nonumber\\
   F&=&\rmd A_{t,s}+A_{t,s}^2=F_s|_{A=A^{g_t^{-1}}}\,,\qquad
   F_s=s\rmd A+s^2A^2
\label{axten}
\end{eqnarray}
in eq.~(\ref{axfour}) as
\begin{equation}
   c_1\int_{T^{2n}}\int_0^1\rmd t\,n\int_0^1\rmd s\,s
   \str\{[\rmd(g_t^{-1}\partial_t g_t) +[A,g_t^{-1}\partial_t
       g_t]]AF_s^{n-1}\}|_{A=A^{g_t^{-1}}}+O(a)\,.
\label{axeleven}
\end{equation}
The integrand is the covariant divergence of the Bardeen-Zumino
current~\cite{Bardeen:1984pm} which supplies a difference between the
consistent anomaly and the covariant anomaly. We note that
eq.~(\ref{axnine}) can be written as
\begin{equation}
   c_1\int_{T^{2n}}\int_0^1\rmd t\int_0^1\rmd
   s\,{\partial\over\partial s} \str(g_t^{-1}\partial_t
   g_tF_s^n)|_{A=A^{g_t^{-1}}}+O(a)\,.
\label{axtwelve}
\end{equation}
Then by using the Bianchi identity~$\rmd F_s+s[A,F_s]=0$, as a sum of
eqs.~(\ref{axeleven}) and~(\ref{axtwelve}), we have
\begin{equation}
   \Gammait_{\rm WZW}[g,U]={c_1\over n+1}\int_{T^{2n}\times I}
   \omega_{2n}^1(g_t^{-1}\rmd_tg_t,A^{g_t^{-1}})+O(a)\,,\qquad
   \rmd_t=\rmd t\,{\partial\over\partial t}\,,
\label{axthirteen}
\end{equation}
where the interval~$I=[0,1]$ stands for the integration region
of~$t$. The consistent anomaly $\omega_{2n}^1$ in this expression is
defined by
\begin{equation}
   \omega_{2n}^1(C,A)=n(n+1)\int_0^1\rmd s\,(1-s)\str[C\rmd(AF_s^{n-1})]\,.
\label{axfourteen}
\end{equation}
Eq.~(\ref{axthirteen}) is the standard expression of the WZW term in
the continuum (i.e., the integrated anomaly) and establishes that our
construction~(\ref{twoxfifty}) has the correct continuum limit. The
following is merely a copy of the standard argument in the
continuum~\cite{Bertlmann:xk}.

Eq.~(\ref{axthirteen}) can be written in a $2n+1$ dimensional form by
using the Chern-Simons form~$\omega_{2n+1}$ as
\begin{equation}
   \Gammait_{\rm WZW}[g,U]={c_1\over n+1}\int_{T^{2n}\times I}
   \omega_{2n+1}(\mathcal{A}^{g_t^{-1}},\mathcal{F}^{g_t^{-1}})+O(a)\,,
\label{axfifteen}
\end{equation}
where the Chern-Simons form is defined by
\begin{equation}
   \omega_{2n+1}(A,F)=(n+1)\int_0^1\rmd s\,\tr(AF_s^n)
\label{axsixteen}
\end{equation}
and $\mathcal{A}^{g_t^{-1}}$ and~$\mathcal{F}^{g_t^{-1}}$ are gauge
fields extended to $2n+1$ dimensions by using the gauge
transformation:
\begin{equation}
   \mathcal{A}^{g_t^{-1}}=g_t^{-1}(\rmd+\rmd_t+A)g_t\,,\qquad
   \mathcal{F}^{g_t^{-1}}=g_t^{-1}Fg_t\,.
\label{axseventeen}
\end{equation}
The gauge transformation law of the Chern-Simons form is
well-known~\cite{Bertlmann:xk}:
\begin{equation}
   \omega_{2n+1}(\mathcal{A}^{g_t^{-1}},\mathcal{F}^{g_t^{-1}})
   =\omega_{2n+1}(A,F)+(\rmd+\rmd_t)\alpha_{2n}
   +\omega_{2n+1}(g_t^{-1}(\rmd+\rmd_t)g_t,0)\,.
\label{axeighteen}
\end{equation}
For lower dimensions, the explicit form of~$\alpha_{2n}$ is given by
\begin{eqnarray}
   \alpha_2(V,A)&=&-\tr(VA),
\nonumber\\
   \alpha_4(V,A)&=&-{1\over2}
   \tr\left[V(A\rmd A+\rmd AA+A^3)-{1\over2}VAVA-V^3A\right],
\label{axnineteen}
\end{eqnarray}
where $V=\rmd gg^{-1}$. For higher dimensions, $\alpha_{2n}$ can
systematically be constructed by using Cartan's homotopy
formula~\cite{Bertlmann:xk}. By substituting eq.~(\ref{axeighteen}) in
eq.~(\ref{axfifteen}) and by noting that the $2n+1$ dimensional
integral of $\omega_{2n+1}(A,F)$ vanishes (because it does not
contain~$\rmd t$), we have
\begin{eqnarray}
   \Gammait_{\rm WZW}[g,U]
   &=&{c_1\over n+1}\int_{T^{2n}}\alpha_{2n}(V=\rmd gg^{-1},A)
\nonumber\\&&
   +\,{(-1)^n(n!)^2c_1\over(2n+1)!}\int_{T^{2n}\times I}
   \tr[g_t^{-1}(\rmd+\rmd_t)g_t]^{2n+1}+O(a)\,.
\label{axtwenty}
\end{eqnarray}
In particular, the classical continuum limit of the Witten term
$\Gammait_{\rm W}[g]=\Gammait_{\rm WZW}[g,1]$ is given~by
\begin{equation}
   \Gammait_{\rm W}[g]
   =-{(-1)^ni^{n+1}n!\over(2\pi)^n(2n+1)!}\int_{T^{2n}\times I}
   \tr[g_t^{-1}(\rmd+\rmd_t)g_t]^{2n+1}+O(a)\,.
\label{axtwentyone}
\end{equation}

\paragraph{Note added in proofs.} The interpolations with respect to the parameter~$s$ in
eqs.~(\ref{twoxthirtynine}) and~(\ref{twoxfortysix}) are not uniquely
defined when ``exceptional configurations'' appear within the
power. For example, $-1$ is an element of $\SU(N)$ with even~$N$ and
$(-1)^s$ is not uniquely defined.  If this situation happens, we
deform the one-parameter family~$g_t(x)$ infinitesimally so that only
non-exceptional configurations appear in the power. We thank
David~H. Adams for a comment on this point.

\end{document}